\documentclass[12pt,draftclsnofoot,onecolumn]{IEEEtran}

\usepackage[noadjust]{cite}
\usepackage{graphicx,color,overpic,psfrag}
\usepackage{amsmath, amssymb}
\usepackage{latexsym}
\usepackage{bm}
\usepackage{amssymb}
\usepackage{cases}
\usepackage{array}
\usepackage{fancyhdr}
\usepackage{setspace}
\usepackage{subfigure}
\usepackage{extarrows}
\usepackage{enumerate}
\usepackage{mathrsfs}

\ifCLASSOPTIONcompsoc
\usepackage[caption=false,font=normalsize,labelfont=sf,textfont=sf]{subfig}
\else
\usepackage[caption=false,font=footnotesize]{subfig}
\fi

\usepackage{url}
\usepackage{algpseudocode}
\usepackage{algorithm}
\usepackage{blkarray}
\usepackage{booktabs}

\usepackage{multirow}
\usepackage{dsfont}
\usepackage{tabularx}
\usepackage[table]{xcolor}

\usepackage{amsfonts}

\usepackage{letltxmacro}

\graphicspath{{figure/}}

\newcommand{\ra}[1]{\renewcommand{\arraystretch}{#1}}
\newcommand{\Pmatrix}[1]{\begin{array}{ll}#1\end{array}}

\newtheorem{theorem}{Theorem}
\newtheorem{lemma}{Lemma}
\newtheorem{remark}{Remark}

\newcommand{\secref}[1]{Section \ref{#1}}

\newcommand{\diag}[1]{\mathsf{diag}\left\{#1\right\}}

\newcommand{\GCN}[2]{\mathcal{CN}\left( #1 , #2\right) }

\newcommand{\vecele}[2]{\left[#1\right]_{#2}}

\newcommand{\thetabs}[2]{{\dnnot{\theta}{bs}}}

\newcommand{\cC}{\mathcal{C}}
\newcommand{\cD}{\mathcal{D}}

\newcommand{\cL}{\mathcal{L}}

\newcommand{\cR}{\mathcal{R}}

\newcommand{\bh}{\mathbf{h}}

\newcommand{\bp}{\mathbf{p}}
\newcommand{\bq}{\mathbf{q}}

\newcommand{\bw}{\mathbf{w}}

\newcommand{\bA}{\mathbf{A}}
\newcommand{\bB}{\mathbf{B}}

\newcommand{\bH}{\mathbf{H}}
\newcommand{\bI}{\mathbf{I}}

\newcommand{\bN}{\mathbf{N}}

\newcommand{\bQ}{\mathbf{Q}}
\newcommand{\bR}{\mathbf{R}}
\newcommand{\bS}{\mathbf{S}}
\newcommand{\bT}{\mathbf{T}}

\newcommand{\bV}{\mathbf{V}}
\newcommand{\bW}{\mathbf{W}}
\newcommand{\bX}{\mathbf{X}}

\newcommand{\bbC}{\mathbb{C}}
\newcommand{\bbR}{\mathbb{R}}

\newcommand{\bLambda}{{\boldsymbol\Lambda}}

\newcommand{\bOmega}{{\boldsymbol\Omega}}
\newcommand{\bomega}{{\boldsymbol\omega}}

\newcommand{\bXi}{{\boldsymbol\Xi}}

\newcommand{\dnnot}[2]{#1_{\mathrm{#2}}}

\allowdisplaybreaks

\begin{document}
	
\title{ Robust Precoding in Massive MIMO: A Deep Learning Approach}
\author{
	Junchao~Shi,~\IEEEmembership{Student~Member,~IEEE},~Wenjin~Wang,~\IEEEmembership{Member,~IEEE},\\
	~Xinping~Yi,~\IEEEmembership{Member,~IEEE},~Xiqi~Gao,~\IEEEmembership{Fellow,~IEEE},\\
	and~Geoffrey~Ye~Li,~\IEEEmembership{Fellow,~IEEE}
	\thanks{
		J. Shi, W. Wang, and X. Q. Gao are with the National Mobile Communications Research Laboratory, Southeast University, Nanjing 210096, China (e-mail: jcshi@seu.edu.cn; wangwj@seu.edu.cn; xqgao@seu.edu.cn).
	}
	\thanks{X. Yi is with the Department of Electrical Engineering and Electronics, University of Liverpool, L69 3BX, United Kingdom (email: xinping.yi@liverpool.ac.uk).
	}
	\thanks{
		G. Y. Li is with the School of Electrical and Computer Engineering, Georgia Institute of Technology, Atlanta, GA 30332, USA (e-mail: liye@ece.gatech.edu).
	}
}
\maketitle
\vspace{-0.3cm} 

\begin{abstract}
In this paper, we consider massive multiple-input-multiple-output (MIMO) communication systems with a uniform planar array (UPA) at the base station (BS) and investigate the downlink precoding with imperfect channel state information (CSI).
By exploiting both instantaneous and statistical CSI, we aim to design precoding vectors to maximize the ergodic rate (e.g., sum rate, minimum rate and etc.) subject to a total transmit power constraint.
To maximize an upper bound of the ergodic rate, we leverage the corresponding Lagrangian formulation and identify the structural characteristics of the optimal precoder as the solution to a generalized eigenvalue problem. 
As such, the high-dimensional precoder design problem turns into a low-dimensional power control problem.
The Lagrange multipliers play a crucial role in determining both precoder directions and power parameters, yet are challenging to be solved directly.
To figure out the Lagrange multipliers, we develop a general framework underpinned by a properly designed neural network that learns directly from CSI.
To further relieve the computational burden, we obtain a low-complexity framework by decomposing the original problem into computationally efficient subproblems with instantaneous and statistical CSI handled separately.
With the off-line pretrained neural network, the online computational complexity of precoding is substantially reduced compared with the existing iterative algorithm while maintaining nearly the same performance.
\end{abstract}
\vspace{-0.3cm} 
\begin{IEEEkeywords}
Robust precoding, solution structure, deep learning, massive MIMO
\end{IEEEkeywords}

\section{Introduction}\label{Sec:Intro}

By deploying a large number of antennas at the base station (BS),  massive multiple-input-multiple-output (MIMO) technique improves spectrum efficiency while serving multiple users as the same time \cite{Erik2014Massive,Marzetta2016Fundamentals,Bruno2016Rate}.
With a huge number of antennas, either in a linear or planar array, the BS can steer the precoding directions accurately to alleviate the interference among users.

Over the past several years, downlink precoder design for massive MIMO has attracted extensive interest \cite{Liang2014Low,Jaehyun2015Multi,Jacobsson2017Quantized}.
In quasi-static and low-mobility scenarios, the available instantaneous channel state information (CSI) at the BS is relatively accurate.
In this situation, linear precoding methods, e.g., regularized zero-forcing (RZF), signal-to-leakage-and-noise ratio (SLNR), and weighted minimum mean-squared error (WMMSE) \cite{Wagner2012Large, Sadek2007Active, Christensen2008Weighted}, can easily achieve multiplexing gain \cite{Caire2010Multiuser}.
Among them, precoder for sum rate maximization can be obtained by iteration in \cite{Bjornson2014Optimal}, which is relatively simple, but iteration still incurs processing delay and is intolerable sometimes.
To address this issue, the recent work \cite{Xia2019A} has used deep learning for downlink beamforming with instantaneous CSI.

The performance of precoders depends on the accuracy of available instantaneous CSI at the transmitter (CSIT) \cite{Kammoun2014Linear}.
Its availability relies on downlink estimation and uplink feedback in a frequency division duplexing system.
Nevertheless, it is extremely difficult to obtain the perfect CSIT in practical systems due to heavy pilot overhead \cite{You2015Pilot} and channel estimation errors \cite{Mi2017Massive}, etc.
Furthermore, for high-mobility networks, relatively short channel coherence time also results in more challenges on CSI acquisition.
In brief, CSIT obsolescence and error often incur serious performance degradation for the precoding methods relying highly on instantaneous CSI.

Even if instantaneous CSI varies with time, statistical CSI usually changes slowly.
Thus, a unified precoding framework can make use of both instantaneous and statistical CSI to adapt the change of the varying communication environment.
The recent work in \cite{Lu2019Robust} has proposed a posteriori channel model to capture both instantaneous and statistical CSI to design robust precoder.
The spatial domain correlation characteristics \cite{Weichselberger2006A} can be further used to address the effects of channel estimation error and channel aging.

While the use of statistical CSI can improve the robustness in precoding design, we must find the corresponding ergodic rate first, which requires to average the instantaneous rate over a large number of channel samples and is challenging.
The iterative algorithm in \cite{Lu2020Robust} can achieve near-optimal performance at the expense of high computational complexity and processing delay.

The recent success of deep learning (DL) in many related areas has motivated its exploration in wireless communications \cite{Yan2017Signal,Timothy2017An,He2018Deep,Hao2018Power,Zhang2019Deep,Ye2019Deep,Zhijin2019Deep}, including channel estimation and prediction, signal detection, resource allocation and etc.
In this paper, we will investigate DL for low-complexity robust precoder design.
The convolutional neural network (CNN) has been applied for feature extraction from CSI \cite{Liang2018An} and for CSI feedback and recovery \cite{Liu2019Exploiting}.
Despite many successful cases in DL for wireless communications \cite{C2019Deep}, it is challenging, if not infeasible, to use DL for precoder design for the high dimensional precoding vectors as the output makes neural networks difficult to be trained.
Thus, it is critical to find a way to convert the high-dimensional precoding problem into a low-dimensional parameter-learning one.

In this paper, we consider the posteriori model that captures both instantaneous and statistical CSI and formulate robust precoding design as an ergodic rate (e.g. sum rate, minimum rate) maximization problem subject to a power constraint.
To make this problem tractable, we transform it into an improved Quality-of-Service (QoS) problem instead of maximizing an upper bound of the ergodic rate, by which the structure of optimal precoding is characterized.
By means of a deep neural network, the proposed structure, can successfully reduce the dimension of the problem and achieve outstanding performance.
In summary, our contributions in this work are three-fold.
\begin{itemize}
	\item By a Lagrangian reformulation, we characterize the structure of optimal precoding vectors, whose direction and power can be associated with the solution to a generalized eigenvalue problem.
	Once the Lagrange multipliers are determined, the precoding vectors can be immediately computed, which transforms the high-dimensional precoding problem into the low-dimensional Lagrangian multiplier computing problem.
	\item To determine the Lagrange multipliers, we use neural networks to learn the mapping from CSI to Lagrange multipliers, and therefore can immediately obtain the precoding vectors.
	\item We develop a low-complexity framework and decompose the original problem into two parts with instantaneous and statistical CSI considered separately.
	Thus, two Lagrange multipliers are computed respectively, followed by a weighted combination.
\end{itemize}
Compared with the existing methods, our general framework significantly reduces the computational complexity while maintaining near-optimal performance.

The rest of this paper is organized as follows.
In \secref{Sec:Model}, we present the posteriori channel and signal model.
In \secref{Sec:Analysis}, we formulate the problem and further investigate the optimal solution structure.
In \secref{Sec:GF}, we develop a general framework for robust precoding based on neural networks.
In \secref{Sec:LF}, we develop a low-complexity framework to further reduce the computational complexity.
Simulation results are presented in \secref{Sec:Simulation} and the paper is concluded in \secref{Sec:Conclusion}.

Some of the notations used in this paper are listed as follows:
\begin{itemize}
	\item Upper and lower case boldface letters denote matrices and column vectors, respectively.
	\item $\bbC^{M\times N}$ ($\bbR^{M\times N}$) denotes the $M\times N$ dimensional complex (real) matrix space, ${\bI}_{N}$ denotes the $N\times N$ identity matrix and the subscript for dimention is sometimes omitted for brevity.
	\item $\odot$ and $\otimes$ denote the Hadamard and Kronecker product of two matrices, respectively.
	\item $\mathbb{E} \left\{  \cdot  \right\}$ denotes the expectation operation, $\triangleq$ denotes the definition, $(\cdot)^{H}$, $(\cdot)^{T}$, and $(\cdot)^{*}$ denote conjugate transpose Hermitian, transpose, and complex conjugate operations, respectively.
	\item $\vecele{\cdot}{i}$ and $\vecele{\cdot}{ij}$ denote the $i$-th element of a vector and the $(i,j)$-th element of a matrix, respectively.
	\item ${\rm tr}(\cdot)$ and ${\rm det}(\cdot)$ represent matrix trace and determinant operations, respectively.
	\item $\sim$ denotes `be distributed as', and $\GCN{\bm \alpha}{\bB}$ denotes the circular symmetric complex Gaussian distribution with mean ${\bm \alpha}$ and covariance $\bB$.
	\item $\diag{\bA}$ denotes the vector along the main diagonal of $\bA$ and the inequality $\bA \succeq {\bf 0}$ means that $\bA$ is Hermitian positive semi-definite.
\end{itemize}
\section{System and Channel Models}\label{Sec:Model}

\begin{figure}[htb]
	\setlength{\abovecaptionskip}{2pt}
	\setlength{\belowcaptionskip}{8pt}
	\centering
	\includegraphics[width=4 in]{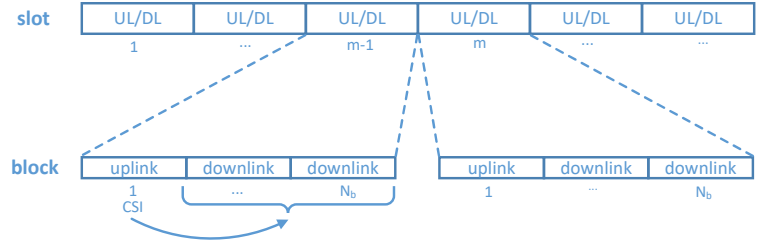}
	\caption{TDD frame with blocks.}\label{Fig:TimeStructure}
\end{figure}

Consider downlink transmission of massive MIMO consisting of one BS and $K$ users.
The BS is equipped with an $M_{v} \times M_{h}$ uniform planar array (UPA), where $M_{v}$ and $M_{h}$ denote the numbers of vertical column and horizontal row, respectively.
Thus, the number of antennas at the BS is $M_t = M_{v} M_{h}$.
Each UE is equipped with a single antenna.
For a time division duplexing (TDD) system, downlink and uplink transmissions are organized into slots, each consisting of $N_b$ blocks.
As can be illustrated in Fig. \ref{Fig:TimeStructure}, in each slot, the blocks can be classified as `uplink', or `downlink' \cite{3GPP38211} for uplink sounding and downlink transmission, respectively.
The first block of each slot contains the uplink sounding signal.
 
\subsection{Channel Model}
The widely-adopted jointly correlated channel model in \cite{Weichselberger2006A} uses the discrete fourier transform (DFT) matrix to represent the spatial sampling matrix.
In this paper, we replace the DFT matrix with the oversampling one to capture the spatial correlation at each subchannel.
Denote $N = N_h N_v$, where $N_v$ and $N_u$ are the vertical and horizontal oversampling factors, respectively.
The spatial sampling matrix can therefore be represented by \cite{Lu2020Robust,Junil2014Bounds,Dawei2014Kronecker}
\begin{align}
\bV_{M_t} = \bV_{M_h} \otimes \bV_{M_v} \in \mathbb{C}^{ M_t \times NM_t},
\end{align}
where the oversampling DFT matrices for the horizontal and vertical directions are respectively given by
\begin{align}
\bV_{M_h} = \frac{1}{\sqrt{M_h}} \Big( e^{ \frac{ -j 2 \pi m n }{N_h M_h} }\Big)_{m=0,\ldots,M_h-1,n=0,\ldots,N_hM_h-1},
\end{align}
and
\begin{align}
\bV_{M_v} &= \frac{1}{\sqrt{M_v}} \Big( e^{ \frac{ -j 2 \pi m n }{N_v M_v} }\Big)_{m=0,\ldots,M_v-1,n=0,\ldots,N_vM_v-1}.
\end{align}

To capture the correlation across different blocks, we utilize the first-order Gauss-Markov process to model the time variation of the channel from one block to another.
It is assumed that the channel keeps unchanged at each block and varies across blocks, so that the precoder is carried once at each block.
The obtained channel estimation at the first block will be used for the current slot.
Thus, by taking into account time correlation, the channel of the $k$-th user at the $n$-th block of the $m$-th slot can be represented by the posteriori model \cite{Lu2020Robust}
\begin{equation}\label{Equ:Posteriori_UPA}
\bh_{k,m,n} = \beta_{k,m,n} \bar \bh_{k,m} + \sqrt{1-\beta_{k,m,n}^2} \bV_{M_t} ({\bf m}_k \odot \bw_{k,m,n}) \in \mathbb{C}^{M_t \times 1},
\end{equation}
where $\bar \bh_{k,m}$ denotes the estimated instantaneous channel, ${\bf m}_k \in \mathbb{C}^{N M_t \times 1}$ is a deterministic vector with nonnegative elements satisfying $\bomega_k = {\bf m}_k \odot {\bf m}_k$, $\bomega_k$ is the channel coupling matrices (CCMs), $\bw_{k,m,n} \in \mathbb{C}^{N M_t \times 1}$ is a complex Gaussian random vector of independent and identically distributed (i.i.d.) entries with zero mean and unit variance, $\beta_{k,m,n} \in [0,1]$ is the time correlation coefficient.
By adjusting $\beta_{k,m,n}$, the posteriori model can leverage channel uncertainties between instantaneous and statistical CSI in various mobile scenarios, e.g., when it tends to be $1$, the channel tends to quasi-static and instantaneous CSI comes to play, and when it tends to be $0$, it corresponds to a high-mobility scenario where only statistical CSI is available.

\subsection{Downlink Transmission}
We now consider the downlink transmission in one block of one slot; therefore, we omit $m$ and $n$ in the subscript hereafter.
Denote $x_{k} \in \mathbb{C}$ the transmitted signal to the $k$-th user.
The received signal of the $k$-th user is given by
\begin{equation}
y_{k} = \bh_{k}^H \bp_{k} x_{k} + \sum\limits_{j \neq k}^K \bh_{k}^H \bp_{j} x_{j} + n_{k}  \in \mathbb{C},
\end{equation}
where $\bp_{k} \in \mathbb{C}^{M_t \times 1}$ is the precoding vector of the $k$-th user, and $n_{k}\sim {\cal C}{\cal N} ( 0,\sigma_n^2 )$ is a complex Gaussian noise.
The ergodic achievable rate of the $k$-th user is given by
\begin{align}
\cR_k = \mathbb E \big\{ \log  (\sigma^2_n + \sum_{i = 1}^{K} \bh_k^H\bp_i\bp_i^H\bh_k  ) \big\} - \mathbb E \big\{ \log  (\sigma^2_n + \sum_{i \neq k}^{K} \bh_k^H\bp_i\bp_i^H\bh_k  ) \big\},
\end{align}
where the precoding vector satisfies the total power constraint $\sum\nolimits_{k=1}^{K} \bp_k^H\bp_k  \le P$.

\section{Optimal Precoding Structure Analysis}\label{Sec:Analysis}
In this section, we formulate the robust precoding problem and characterize the structure of optimal precoding vectors by maximizing an upper bound of the egrodic rate.
\subsection{Problem Formulation}
The objective is to design precoding vectors $\bp_{1},...,\bp_{K}$ that maximize an utility function of ergodic rate as follows
\begin{align}\label{Equ:MaxFun}
{\bf P1:}\;\;\;\;&\mathop {\max }\limits_{\bp_{1},...,\bp_{K}} f(\cR_1,\ldots,\cR_K), \nonumber\\
&\;\;\;\;\; {\rm{s.t.}} \;\;   \sum\limits_{k=1}^{K} \bp_k^H\bp_k  \le P, \;\;\;\; k = 1,\ldots,K,
\end{align}
where $f(\cR_1,\ldots,\cR_K)$ can be any function, e.g, sum rate and minimum rate, and $P$ denotes the total power budget.

This optimization problem involves high-dimensional variables and the objective function is non-convex in general.
As a result, the exact solution is intractable.
Although there exist various approximation methods, the high dimensionality of the optimization variables usually demands high computation to achieve optimal performance.
For example, the iterative approach in \cite{Lu2020Robust} can nearly achieve the maximum sum rate.
To reduce computational complexity, we aim to explore a solution structure of the precoding to transform the high-dimensional optimization problem to a low-dimensional one.

\subsection{Problem Transformation}
First, we introduce the following lemma to bridge our formulation to a QoS problem, proved in Appendix \ref{Proof:EquivalentProblem}.
\begin{lemma}\label{Lemma:EquivalentProblem}
	Denote $\cR_1^{\Diamond},\ldots,\cR_K^{\Diamond}$ the ergodic rates achieved by a solution (referred to as $\bf {S1}$) of $\bf P1$.
	The optimal solution (referred to as $\bf {S2}$) of the following QoS problem achieves the same ergodic rates as $\bf {S1}$ but with lower or equal total power. 
	\begin{align}\label{Equ:Minpower}
	{\bf P2:}\;\;\;\;&\mathop {\min }\limits_{\bp_{1},...,\bp_{K}} \sum\limits_{k=1}^{K} \bp_k^H\bp_k, \nonumber\\
	&\;\;\;\;\; {\rm{s.t.}} \;\; \cR_k \ge \cR_k^\Diamond, \;\;\;\; k = 1,\ldots,K.
	\end{align}
	When $\bf {S1}$ is the global optimal, $\bf {S2}$ is equivalent to $\bf {S1}$, i.e., achieves the same ergodic rates and total power.
\end{lemma}

Lemma \ref{Lemma:EquivalentProblem} indicates $\bf P2$ can improve or maintain any solution of $\bf P1$. 
By converting the problem into such a QoS problem, the ergodic rate of each user can be decoupled to the constraints.
As these optimal rates are demanded, this reformulation, while not directly help solve $\bf P1$, can help understand the structure of the optimal precoding vectors.

\subsection{Optimal Solution Structure}
Noting that the constraint $\cR_k \ge \cR_k^\Diamond$ always holds in the case of $\cR_k^\Diamond = 0$ and clearly the corresponding solution is $\bp_k^\Diamond = {\bf 0}$, we conclude that the users with zero-rate can be eliminated from $\bf P2$.
Consequently, we here assume $\cR_k^\Diamond >0$ without loss of generality.

As there exists no closed form of the ergodic rate, direct optimization of $\bf P2$ is intractable.
Thus, we employ the following upper bound
\begin{align}
\cR_k \le \cR_k^{ub} \triangleq\log \big (\sigma_n^2 + \sum_{i = 1}^{K} \mathbb E \{ \bh_k^H\bp_i\bp_i^H\bh_k \} \big ) - \log \big (\sigma_n^2 + \sum_{i \neq k}^{K} \mathbb E \{ \bh_k^H\bp_i\bp_i^H\bh_k \} \big ),
\end{align}
which is due to Jensen's inequality to make the problem more tractable.
By doing so, the constraints can be transformed into the following tractable quadratic form
\begin{align}\label{Equ:Constraint}
&\cR_k^{ub} \ge {\cR_k^{ub}}^\Diamond \Longleftrightarrow {\rm SINR}_k \ge \gamma_k \Longleftrightarrow \cC_k \le 0, \forall k,
\end{align}
where the signal-to-interference-plus-noise-ratio (SINR) of the $k$-th user is given by
\begin{align}
{\rm SINR}_k = \frac{\bp_k^H \bR_k \bp_k} { \sigma_n^2 + \sum\nolimits_{i \neq k}^K \bp_i^H \bR_k \bp_i},
\end{align}
$\gamma_k = 2^{{\cR_k^{ub}}^\Diamond} - 1$ can be regarded as the SINR achieved by $\bf S1$, the constraint function is defined as
\begin{equation}
\cC_k \triangleq 1 + \frac{1}{\sigma_n^2}\sum_{i \neq k}^{K}  \bp_i^H\bR_k\bp_i - \frac{1}{\sigma_n^2\gamma_k} \bp_k^H\bR_k\bp_k,
\end{equation}
and $\bR_k = \mathbb E \{ \bh_k\bh_k^H \} \in \bbC^{M_t \times M_t}$.
The optimization problem can be reformulated as
\begin{align}\label{Equ:Minpower_Sum_Final}
{\bf P3:}\;\;\;\;&\mathop {\min }\limits_{\bp_{1},...,\bp_{K}} \sum\limits_{k=1}^{K} \bp_k^H\bp_k, \nonumber \\
&\;\;\;\;\;{\rm{s.t.}} \;\; \cC_k \le 0, \;\;\;\; k = 1,\ldots,K.
\end{align}
The appropriate transformation lends itself to the analysis of the following solution structure.

The Lagrangian of $\bf P3$ can be expressed as
\begin{align}
\cL_\cR = \sum_{k=1}^{K} \bp_k^H\bp_k + \sum_{k=1}^{K} \mu_k \cC_k,
\end{align}
where $\mu_k$ is the Lagrange multiplier.
The derivative of $\cL_\cR$ can be written as
\begin{equation}
\frac{\partial \cL_\cR}{\partial \bp_k}= \bp_k + \sum_{i \neq k}^{K} \frac{\mu_i}{\sigma_n^2} \bR_i\bp_k - \frac{\mu_k}{\sigma_n^2\gamma_k} \bR_k \bp_k.
\end{equation}
Denote that ${\bm \mu} = (\mu_1,\ldots,\mu_K)^T \in \mathbb{C}^{K \times 1}$.
The optimal solution of $\bf P3$ should satisfy the following Karush-Kuhn-Tucker (KKT) conditions \cite{Boyd2004Convex}
\begin{align}
\frac{\partial \cL_\cR}{\partial \bp_k} ({\bm \mu},\bp_k) = 0, &\;\;\;\; k = 1,\ldots,K, \label{Equ:KKT_D}\\
\mu_k \cC_k = 0, &\;\;\;\; k = 1,\ldots,K, \label{Equ:KKT_C}\\
\mu_k \ge 0, &\;\;\;\; k = 1,\ldots,K.
\end{align}
Denote $\bp_k = \sqrt{\rho_k} {\underline \bp}_k$, where $\rho_k$ is the power allocated on the $k$-th user, ${\underline \bp}_k$ is the normalized precoding vector satisfying ${\underline \bp}_k^H {\underline \bp}_k = 1$.
According to the above derivation, we can investigate the precoding characteristics in the following.

\subsubsection{Generalized Eigen Domain Precoding}
According to (\ref{Equ:KKT_D}), we can obtain
\begin{align}\label{Equ:GEVP}
\mu_k \bR_k \underline\bp_k = \gamma_k \Big (\sigma_n^2 \bI + \sum_{i \neq k}^{K} \mu_i \bR_i \Big ) \underline\bp_k.
\end{align}
This is a well-known generalized eigenvalue problem.
According to (\ref{Equ:Posteriori_UPA}), the covariance matrices can be computed by
\begin{align}
\bR_k = \beta_k^2 \bar\bh_k \bar\bh_k^H + (1-\beta_k^2) \bV_{M_t} \bLambda_k \bV_{M_t}^H,
\end{align}
where $\bLambda_k \in \bbC^{NM_t \times NM_t}$ is diagonal with $[\bLambda_k]_{ii} = [\bomega_k]_i,\forall i$.
The computation of $\mu_k$ will be discussed in next section.
Denote
\begin{align}
\bS_k = \mu_k \bR_k,
\end{align}
and
\begin{align}
\bN_k = \sigma_n^2 \bI + \sum_{i \neq k}^{K} \mu_i \bR_i,
\end{align}
then $\underline\bp_k$ is the generalized eigenvector with respect to generalized eigenvalue $\gamma_k$ of matrix pair $(\bS_k,\bN_k)$.
Although $\gamma_k$'s are unknown, it is not necessarily to compute them in advance due to the following theorem, proved in Appendix \ref{Proof:MaxEigvalue}.
\begin{theorem}\label{Theorem:MaxEigvalue}
	The optimal solution of $\bf P3$ is the generalized eigenvector with respect to the maximum generalized eigenvalue of matrix pair $(\bS_k,\bN_k)$, i.e.,
	\begin{subequations}\label{Equ:MaxEigvalue}
	\begin{align}
	\underline\bp_k &= \max. {\rm generalized \; eigenvector} (\bS_k,\bN_k), \\
	\gamma_k &= \max. {\rm generalized \; eigenvalue} (\bS_k,\bN_k).
	\end{align}
	\end{subequations}
\end{theorem}

Theorem \ref{Theorem:MaxEigvalue} indicates that once the Lagrange multipliers are determined, the 
precoding direction $\underline \bp_k$ and the parameter $\gamma_k$ can be computed immediately.
The $\gamma_k$'s play a crucial role in computing the precoding powers as discussed in Section \ref{Subsubsec:GEDPC}.

The precoder direction determined by the upper bound of ergodic rate also applies to the maximizing SLNR case, i.e., the weighted SLNR (WSLNR) precoder
\begin{align}
\;\;\;\;\;&\mathop {\max }\limits_{\underline \bp_{k}}{\rm WSLNR}_k = \frac{\mu_k \underline\bp_k^H \bR_k \underline\bp_k} { \sigma_n^2 + \sum\nolimits_{i \neq k}^K \mu_i\underline\bp_k^H \bR_i \underline\bp_k}, \nonumber \\
&\;\;\;\;\;{\rm{s.t.}} \;\; \underline\bp_k^H \underline\bp_k = 1, \;\;\;\; k = 1,\ldots,K.
\end{align}
The key is the introduction of the Lagrange multipliers, which is conducive to reduce the dimension of the problem.
As the optimal Lagrange multipliers are implicit, we propose to compute them by deep neural networks in Section \ref{Subsec:LMNN}.

It is worth pointing out that the structure in \cite{Bjornson2014Optimal} is dedicated to the vector channel with the rank of covariance matrix being 1.
Our proposed structure covers the general case with arbitrary rank.
Actually, the structure in \cite{Bjornson2014Optimal} can be regarded as a special case of (\ref{Equ:MaxEigvalue}), so are some other existing methods.
This implies the universality of the proposed structure.
Below we give the brief analyses.
\begin{remark}
	The SLNR of the $k$-th user can be expressed as
	\begin{align}
	{\rm SLNR}_k = \frac{\underline\bp_k^H \bR_k \underline\bp_k} { \sigma_n^2 + \sum\nolimits_{i \neq k}^K \underline\bp_k^H \bR_i \underline\bp_k}.
	\end{align}
	Accordingly, from \cite{Sadek2007Active}, the SLNR precoder is given by 
	\begin{equation}\label{Equ:SLNR}
	\underline\bp_k = \max. {\rm generalized \; eigenvector} (\bR_k, \sigma_n^2 \bI + \sum_{i \neq k}^{K} \bR_i ).
	\end{equation}
	If we set $\mu_k=1, \forall k$, (\ref{Equ:MaxEigvalue}) boils down to (\ref{Equ:SLNR}), which is the optimal precoder that maximizes SLNR.
	In general, the SLNR precoder does not sufficiently lead to maximum sum rate while the introduction of the Lagrange multipliers improves the resulting sum rate to the maximum.
\end{remark}
\begin{remark}
	When $\beta_k = 1, \forall k$, (\ref{Equ:GEVP}) turns to the structure in \cite{Bjornson2014Optimal}
	\begin{align}
	\underline\bp_k = \xi_k \mu_k \big( \sigma_n^2 \bI + \sum\nolimits_{i = 1}^K \mu_i \bar\bh_i \bar\bh_i^H \big)^{-1}  {\bar\bh}_k,
	\end{align}
	where $\xi_k = (1 + \frac{1}{\gamma_k}) {\bar\bh}_k^H \underline\bp_k$.
	If we set $\mu_k=\frac{1}{K}, \forall k$, it becomes the RZF precoder.
	In this sense, (\ref{Equ:MaxEigvalue}) can be regarded as the weighted RZF precoder.
	By introducing the Lagrange multipliers, the performance of the RZF precoder can be immediately improved to WMMSE.
\end{remark}
\begin{remark}
	When $\beta_k = 0, \forall k$, we have $\bR_k = \bV_{M_t} \bLambda_k \bV_{M_t}^H$. If we set $N_h = N_v = 1$, then $\bV_{M_t}^H \bV_{M_t} = \bI_{M_{t}}$, (\ref{Equ:GEVP}) becomes
	\begin{equation}
	\mu_k \bLambda_k \underline \bq_k = \gamma_k \big( \sigma_n \bI + \sum_{i \neq k} \mu_i \bLambda_i\big) \underline \bq_k
	\Longleftrightarrow \bXi_k \underline \bq_k = \gamma_k \underline \bq_k, 
	\end{equation}
	where $\underline \bq_k = \bV_{M_t}^H \underline\bp_k$ and $\bXi_k \in \bbC^{NM_t \times NM_t}$ is diagonal and with
	\begin{align}
		[\bXi_k]_{ii} = [\mu_k \big( \sigma_n \bI + \sum_{i \neq k} \mu_i \bLambda_i\big)^{-1} \bLambda_k]_{ii}, \forall i.
	\end{align}
	Denote $m_k = \arg\max\limits_{i} [\bXi_k]_{ii}$ the index of the maximum diagonal element, we have
	\begin{align}\label{Equ:MaxBeam}
	[\underline \bq_k]_i = \left\{ \Pmatrix{1, & \text{if $i = m_k$,}\\0, & \text{otherwise.}} \right.
	\end{align}
	As such, the precoding vector $\bp_k = \bV_{M_t} {\underline \bq}_k$ is the $m_k$-th column of $\bV_{M_t}$.
	In this sense, (\ref{Equ:MaxEigvalue}) can be regarded as an extension of beam division multiple access (BDMA) transmission \cite{Chen2015Beam} and the introduction of the Lagrange multipliers provides a criterion of beam selection to maximize the sum rate.
\end{remark}

Noting that the generalized eigenvector only contains the direction information of the precoding vectors.
The SLNR precoder usually considers equal power allocation, which is generally not optimal.
In fact, the power can be computed by another KKT condition, which will be discussed below.

\subsubsection{Generalized Eigen Domain Power Control}\label{Subsubsec:GEDPC}
According to (\ref{Equ:KKT_C}), we have $\mu_k = 0$ or $\cC_k = 0$.
It can be verified that $\mu_k \ne 0$.
If otherwise $\mu_k=0$, substitute it into (\ref{Equ:GEVP}) and we have $\bp_k = {\bf 0}$, which contradicts the fact that $\cC_k < 0$.
This can also be explained from another point of view.
As has been proved in Appendix \ref{Proof:EquivalentProblem}, the constraint of the optimal solution in $\bf P2$ takes the equal sign, i.e., $\cC_k = 0$.
Thus, we have
\begin{equation}\label{Equ:KKT_C_1}
\sigma_n^2 + \sum_{i \neq k}^{K} \rho_i \underline\bp_i^H \bR_k \underline\bp_i - \frac{\rho_k}{\gamma_k} \underline\bp_k^H \bR_k \underline\bp_k= 0.
\end{equation}
Denote
\begin{align}\label{Equ:T}
t_{ki} = 
\begin{cases}
\frac{1}{\gamma_k}\underline\bp_i^H \bR_k \underline\bp_i, & k = i, \\
-\underline\bp_i^H \bR_k \underline\bp_i,                  & k \neq i.
\end{cases}
\end{align}
We can rewritten (\ref{Equ:KKT_C_1}) as
\begin{align}\label{Equ:KKT_C_2}
\sum_{i =1}^{K} t_{ki} \rho_i = \sigma_n^2, \;\;\;\; k=1,\ldots,K,
\end{align}
the matrix form of which is $\bT {\bm \rho} = \sigma_n^2 {\bf 1}_{K \times 1}$, where $[\bT]_{ki} = t_{ki}$ and ${\bm \rho} = (\rho_1,\ldots,\rho_K)^T$.
To compute the power vector, we first propose the following lemma, proved in Appendix \ref{Proof:Non-singular}.
\begin{lemma}\label{Lemma:Non-singular}
The matrix $\bT$ is non-singular.
\end{lemma}

Thus, the power vector can be computed by
\begin{align}\label{Equ:Cal_rho}
{\bm \rho} = \sigma_n^2 \bT^{-1} {\bf 1}_{K \times 1}.
\end{align}
It is worth mentioning that the precoding vectors computed by the solution structure, i.e., (\ref{Equ:MaxEigvalue}) and (\ref{Equ:Cal_rho}), always satisfy the total power constraint as the optimal Lagrange multipliers satisfy (proved in Appendix \ref{Proof:MaxEigvalue})
\begin{align}
\sum\nolimits_{k=1}^{K} \rho_k = \sum\nolimits_{k=1}^{K} \mu_k \le P.
\end{align}
The precoding power cannot be determined directly as the $\gamma_k$'s are unknown.
However, it can be connected with the Lagrange multipliers thanks to Theorem \ref{Theorem:MaxEigvalue}.
Beyond the precoding direction, the Lagrange multipliers also determine the $\gamma_k$'s, which further determine the precoding power.

\section{Robust Precoding Based on Neural Networks}\label{Sec:GF}
\begin{figure}[htb]
	\setlength{\abovecaptionskip}{2pt}
	\setlength{\belowcaptionskip}{8pt}
	\centering
	\includegraphics[width=3.6 in]{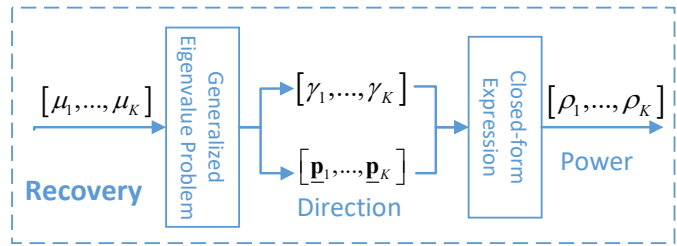}
	\caption{Recovery of the precoding vectors from Lagrange multipliers.}\label{Fig:Recovery}
\end{figure}

Based on the previous analysis, we conclude that the precoding vectors can be recovered losslessly by the Lagrange multipliers.
The diagram of recovery is shown in Fig. \ref{Fig:Recovery}.
The precoding direction can be computed by solving the generalized eigenvalue problem in (\ref{Equ:GEVP}) and the precoding power can be further computed by the closed-form expression in (\ref{Equ:Cal_rho}).
As such, the high-dimensional computation of the precoding vectors turns into low-dimensional Lagrange multipliers, i.e., the key to downlink precoding.
Learning directly the precoding vectors is complicated and difficult to train due to the high dimension of precoding vectors.
However, learning the Lagrange multipliers has no such limitation as the dimension has been much reduced.
In this section, we will propose a general framework for robust precoding by taking advantage of this optimal solution structure, where the Lagrange multipliers are computed by a well trained neural network.

\subsection{Framework Structure}\label{Subsec:GFS}

The following theorem, proved in Appendix \ref{Proof:LMO}, provides the physical meaning of the Lagrange multipliers.
\begin{theorem}\label{Theorem:LMO}
	The optimal solution of the following Lagrange multipliers optimization problem is
	the optimal Lagrange multipliers of $\bf P3$ when $\bf S1$ is global optimal.
	\begin{align}\label{Equ:LMO}
	{\bf P4:}\;\;\;\;&\mathop {\max }\limits_{\mu_{1},...,\mu_{K}} f({\check \cR}_1,\ldots,{\check \cR}_K), \nonumber \\
	&\;\;\;\;\;{\rm{s.t.}} \;\; \sum_{k=1}^K \mu_k \le P,
	\end{align}
	where ${\check \cR}_k = \log \Big(1 + \rho \big( \bN_k^{-1} \bS_k\big)  \Big)$ and $\rho(\cdot)$ denotes the function of the maximum eigenvalue.
\end{theorem}
\begin{remark}
	If we set $\beta_k = 1, \forall k$, as the rank of matrix $\bN_k^{-1} \bS_k$ is $1$, we have
	\begin{equation}
	\check \cR_k = \log\det(\sigma_n^2 \bI +  \sum_{i=1}^{K} \mu_i \bR_i) - \log\det(\sigma_n^2 \bI + \sum_{i \neq k}^{K} \mu_i \bR_i).
	\end{equation}
	As such, the Lagrange multipliers can be regarded as the uplink power parameters and ${\bf P4}$ can be regarded as the power allocation.
	For sum rate maximization, it can be solved by the WMMSE approach \cite{Qingjiang2011An}.
\end{remark}

\begin{figure}[htb]
	\setlength{\abovecaptionskip}{2pt}
	\setlength{\belowcaptionskip}{8pt}
	\centering
	\includegraphics[width=4 in]{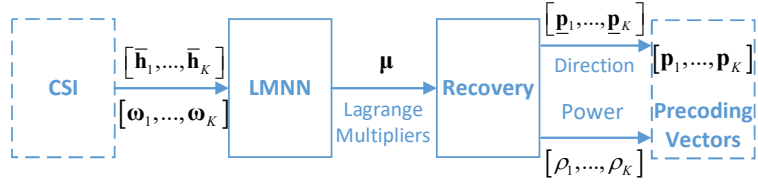}
	\caption{General Framework for Robust Precoding.}\label{Fig:GF}
\end{figure}

However, for the general case, there is no mathematical method available in the literature to solve $\bf P4$.
Thus, we utilize deep learning for this troublesome problem, i.e., the Lagrange multipliers neural network (LMNN).
As shown in Fig. \ref{Fig:GF}, the general framework for robust precoding can be decomposed into three parts:
\begin{enumerate}[i)]
	\item Learn the optimal Lagrange multipliers from the obtained channel matrices;
	\item Compute precoding direction by solving a generalized eigenvalue problem;
	\item Compute precoding power by a closed-form expression in (\ref{Equ:Cal_rho}).
\end{enumerate}

The corresponding algorithm is summarized in Algorithm \ref{Alg:GF}.
Noting that $\bp_k = {\bf 0}$ if $\mu_k = 0$, as there exist slight errors of the neural network, we delete the $k$-th user if $\mu_k \le \epsilon$, where $\epsilon$ is a preset threshold.
\begin{algorithm}[htb]
	\caption{General Framework for Robust Precoding }
	\label{Alg:GF}
	\footnotesize
	\begin{algorithmic}[1]
		\Require The channel matrices $\bar \bh_k$ and $\bomega_k, k=1,\ldots,K$, the noise variance $\sigma_n$ and total power constraint $P$
		\Ensure The precoding matrices $\bp_k, k=1,\ldots,K$
		
		\State Compute the corresponding parameters $\beta_k, k=1,\ldots,K$.
		
		\State Compute the corresponding Lagrange multipliers $\mu_k,k=1,\ldots,K$ and delete users with $\mu_k \le \epsilon$.
		\State Compute the normalized precoding vector $\underline\bp_k$ and the parameter $\gamma_k, k=1,\ldots,K$ by (\ref{Equ:MaxEigvalue}).
		\State Compute the power allocated on the users $\rho_k, k=1,\ldots,K$ by (\ref{Equ:Cal_rho}).
		\State Compute the precoding vectors $\bp_k = \sqrt{\rho_k} \underline\bp_k, k=1,\ldots,K$.
		
	\end{algorithmic}
\end{algorithm}

\subsection{Lagrange Multipliers Neural Network}\label{Subsec:LMNN}
The objective of LMNN is to approximate the Lagrange multipliers from channel matrices.
According to the posteriori model, denote
\begin{align}
	&\bar\bH_{\bm \beta} = [\beta_1\bar\bh_1,\ldots,\beta_K\bar\bh_K]^H \in {\mathbb C}^{K \times M_t},\\
	&\bOmega_{\bm \beta} = [(1-\beta_1^2)\bomega_1,\ldots,(1-\beta_k^2)\bomega_K]^H \in {\mathbb C}^{K \times NM_t},
\end{align}
as the input of the neural network.
Generally, $\bomega_k$ is sparse as $\bV_{M_t}$ is constructed from the oversampling DFT matrix and the CSI contains the original two-dimensional information.
Thus, we utilize CNN to learn the Lagrange multipliers.
The convolutional neural network is composed of several convolution modules, a flatten layer and several fully-connected layers.
Each convolution modules consists of a convolutional layer, an activation function and a pooling layer.
The convolutional layer performs convolutions on the input to extract the feature.
Besides, the widely-adopted rectified linear unit (ReLU) \cite{Nair2010Rectified} (i.e., $h(x) = \max (0,x)$) is chosen as an activation function, which removes negative values to increase nonlinearity and the max-pooling \cite{Ian2016Deep} is chosen for down-sampling.
Next, the flatten layer transforms the feature into a suitable form (i.e., a vector) for the next layers.
Finally, the fully-connected layers accomplish the advanced reasoning by matrix multiplications, where the activation function is also chosen as ReLU.
The Lagrange multipliers are also related to the total power constraint $P$ and noise covariance $\sigma_n^2$, which determines the signal-to-noise ratio (SNR) at the transmitter
\begin{align}
\nu = 10\log \frac{P}{\sigma_n^2}.
\end{align}
The SNR can be included in the channel matrices. However, it will cause great fluctuations in the order of magnitude of the input value under samples with different SNRs.

\begin{figure*}[htb]
	\setlength{\abovecaptionskip}{2pt}
	\setlength{\belowcaptionskip}{8pt}
	\centering
	\includegraphics[width=5.6 in]{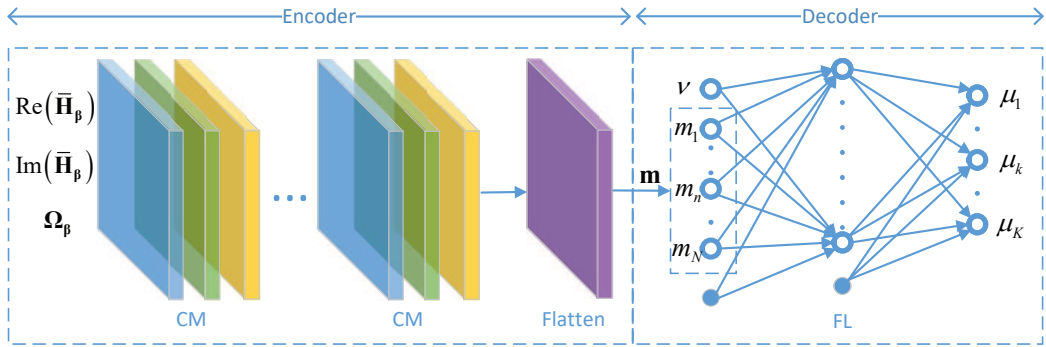}
	\caption{Lagrange Multipliers Neural Network.}\label{Fig:LMNN}
\end{figure*}
As such, we construct the LMNN consisting of a CNN and a fully-connected neural network (FNN), as shown in Fig. \ref{Fig:LMNN}.
The former encodes the channel matrices as the implicit feature and the latter decodes the feature with SNRs as the Lagrange multipliers.
The channel matrix, $\bar \bH_{\bm \beta}$, is divided into the real and imaginary parts.
Backed by the universal approximation theorem of FNN \cite{Cybenko1989Approximation,Hornik1991Approximation} and CNN \cite{Zhou2019Universality}, the LMNN can approximate arbitrary continuous function with arbitrary accuracy as long as the number of neurons is sufficiently large and the depth of the neural network is large enough.
The Lagrange multipliers learning can be decomposed into two steps:
\begin{enumerate}[1)]
	\item \textit{Encoder}: Several convolution modules to encode the CSI as hidden layer feature ${\bm m} = f_{en} (\bar \bH_{\bm \beta}, \bOmega_{\bm \beta}; \bw_{en})$, where $\bw_{en}$ denotes the weight vector of the encoder.
	\item \textit{Decoder}: Several fully-connected layers to decode the hidden layer feature $\bm m$ and the SNR $\nu$ as the Lagrange multipliers ${\bm \mu} = f_{de} (\nu, {\bm m}; \bw_{de})$, where $\bw_{de}$ denotes the weights vector of the decoder.
\end{enumerate}
Thus, the function of LMNN can be written in the form
\begin{align}
{\bm \mu} = f_{\bm \mu} (\bar \bH_{\bm \beta}, \bOmega_{\bm \beta}, \nu; \bw),
\end{align}
where the set of all weight and bias parameters have been grouped together into a vector $\bw$.

\subsection{Dataset Generation and Neural Network Training}
It has been proved that the precoding vectors can be computed by Lagrange multipliers, and interestingly vice versa.
Thus, given the channel matrices, we propose to compute the Lagrange multipliers from precoding vectors by the existing iterative method.
Left-multiplied by $\underline\bp_k^H$, (\ref{Equ:GEVP}) becomes 
\begin{align}
\frac{1}{\gamma_k} \underline\bp_k^H \bR_k \underline\bp_k \mu_k - \sum_{i \neq k}^{K}  \underline\bp_k^H \bR_i \underline\bp_k \mu_i = \sigma_n^2. \label{Equ:GEVP_2}
\end{align}
We can rewritten (\ref{Equ:GEVP_2}) as
\begin{align}\label{Equ:KKT_DataSet}
\sum_{i =1}^{K} t_{ik} \mu_i = \sigma_n^2, \;\;\;\; k=1,\ldots,K,
\end{align}
the matrix form of which is $\bT^H {\bm \mu} = \sigma_n^2 {\bf 1}_{K \times 1}$.
As matrix $\bT$ is non-singular, we can compute the Lagrange multipliers vector by
\begin{align}\label{Equ:Cal_mu}
{\bm \mu} = \sigma_n^2 (\bT^{-1})^H {\bf 1}_{K \times 1}.
\end{align}

In this paper, we consider the weighted sum rate maximization as an example
\begin{align}
f(\cR_1,\ldots,\cR_K) = \cR_{sum} = \sum\nolimits_{k=1}^K w_k \cR_k,
\end{align}
where $w_k$ are real non-negative weights for the balance of fairness between users.
The precoding vectors can be computed by the following iterative equations \cite{Lu2020Robust}
\begin{subequations}\label{Equ:Iterative}
	\begin{align}
	&\mu^{t} \gets \sum_{k=1}^{K} \operatorname{tr}\left(\left(\bp_{k}^{t}\right)^{H}\left(\mathbf{A}_{k}^{t}-\mathbf{B}^{t}\right) \bp_{k}^{t}\right), \\
	&\bp_{k}^{t+1} \gets \left(\bB^{t} + \mu^{t} \bI_{M_{t}}\right)^{-1} \mathbf{A}^{t}_{k} \bp_{k}^{t},
	\end{align}
\end{subequations}
where $t$ denotes the number of iterations, $\bA_{k}= w_k (\sigma_n^2 + \sum_{i \neq k}^{K} \bp_i^H\bR_k\bp_i)^{-1} \bR_k$ and $\bB=\sum_{k=1}^{K}\big( \bA_{k} - w_k (\sigma_n^2 + \sum_{i = 1}^{K} \bp_i^H\bR_k\bp_i)^{-1}\bR_k \big)$.

\begin{algorithm}[!t]
	\caption{Dataset Generation}
	\label{Alg:Dataset}
	\footnotesize
	\begin{algorithmic}[1]
		\Require The number of data samples $N_\cD$
		\Ensure The dataset $\cD$
		
		\State Initialize $i = 1$.
		\While{$i<N_\cD$}
		\State Generate the channel matrices $\bar \bh_k^{(i)}$ and $\bomega_k^{(i)}, k=1,\ldots,K$, the noise variance $\sigma_n^{(i)}$ and total power constraint $P^{(i)}$, compute the coefficient $\beta_k^{(i)},k=1,\ldots,K$ and the SNR $\nu^{(i)}$.
		
		\State Solve the problem (\ref{Equ:MaxFun}) by the iterative approach in (\ref{Equ:Iterative}), compute the precoding vectors $\bp_k^{(i)}$ and the corresponding parameter $\gamma_k^{(i)},k=1,\ldots,K$.
		
		\State Construct the matrix $\bT^{(i)}$ by (\ref{Equ:T}) and compute the corresponding Lagrange multipliers $\mu_k^{(i)},k=1,\ldots,K$ by (\ref{Equ:Cal_mu}).
		
		\State Group $\beta_k^{(i)}$, $\bar \bh_k^{(i)}$, $\bomega_k^{(i)}$, $\nu^{(i)}$ and $\mu_k^{(i)},k=1,\ldots,K$ as the $i$-th sample.
		
		\State Set $i=i+1$.	
		\EndWhile
	\end{algorithmic}
\end{algorithm}
The dataset generation is illustrated in Algorithm \ref{Alg:Dataset}.
As the training is off-line, the precoding vectors can be computed by the high-performance iterative approach without considering much complexity.
In such a case, a sufficiently large enough number of iterations can be set until convergence.
Furthermore, we can select multiple initial values to iterate and choose the best one to avoid some bad local optimal solutions.

Given the training set $\cD$ generated by Algorithm \ref{Alg:Dataset}, the objective is to minimize the loss function
\begin{align}
\cL_\cD = \frac{1}{N_\cD} \sum_{i=1}^{N_\cD} \big\| {{\bm \mu}^{(i)} - \hat{\bm \mu}^{(i)}} \big\|^2,
\end{align}
where $\hat{\bm \mu}^{(i)}$ is the predicted results of the $i$-th sample.
In the training progress, the procedure of dropout \cite{Srivastava2014Dropout} is utilized to avoid over-fitting.
Finally, we employ the widely-used adaptive moment estimation (ADAM) algorithm \cite{Kingma2015Adam} to train the neural network and weights vector $\bw$ can be obtained.

\section{Low-complexity Weighting Framework}\label{Sec:LF}

The proposed general precoding framework based on the neural network has achieved near-optimal performance and the complexity has been significantly reduced compared with the existing iterative algorithm.
However, further simplified computation is desired to be applied in a real-time system.
To this end, we further propose a low-complexity framework in this section.

\subsection{Weighting Strategy for Robust Precoding}
As can be seen in Fig. \ref{Fig:GF}, the complexity is mainly in the following three parts:
\begin{enumerate}[1)]
	\item The neural network for the Lagrange multipliers;
	\item The generalized eigenvalue problem for the precoding direction;
	\item The computation of the precoding power (including the construction of matrix $\bT$).
\end{enumerate}

When only instantaneous CSI is available, the rank of the correlation matrix is one.
Thus, the computational complexity can be much simplified by utilizing mathematical manipulation (e.g., matrix inversion lemma).
When only statistical CSI is used, once computation is required as it remains unchanged for the whole period of time-frequency resources.
Thus, it is an efficient strategy to decompose the general framework into instantaneous and statistical parts.
As the Lagrange multipliers should still satisfy $\sum_{k=1}^K {\mu}_k = P$, we compute the Lagrange multipliers as
\begin{align}\label{Equ:W-mu}
{\mu}_k = \beta_k^2 [{\bm \mu}_{\bh}]_k + (1-\beta_k^2) [{\bm \mu}_\bomega]_k,
\end{align}
where $\bm \mu_\bh$ and $\bm \mu_\bomega$ denote the Lagrange multipliers of the two extremes, respectively.
As the construction of matrix $\bT$ is also time-consuming, we weight the powers with the same strategy.
The precoding power can be computed as
\begin{align}\label{Equ:W-rho}
{\rho}_k = \beta_k^2 [{\bm \rho}_\bh]_k + (1-\beta_k^2) [{\bm \rho}_\bomega]_k,
\end{align}
where $\bm \rho_\bh$ and $\bm \rho_\bomega$ denote the power of the two extremes.

\begin{figure}[htb]
	\setlength{\abovecaptionskip}{2pt}
	\setlength{\belowcaptionskip}{8pt}
	\centering
	\includegraphics[width=4 in]{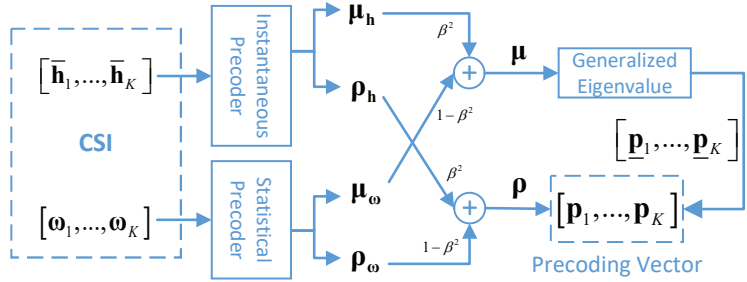}
	\caption{Low-complexity Framework for Robust Precoding.}\label{Fig:LF}
\end{figure}
Denote ${\bm \beta} = [\beta_1,\ldots,\beta_K]$, the low-complexity framework is shown in Fig. \ref{Fig:LF}.
As the Lagrange multipliers and the precoding power can be computed efficiently by the weighting strategy, now we focus on the efficient computation of generalized eigenvalue problem.
It can be solved by transforming it into a standard eigenvalue problem with the operation of matrix inversion.
However, due to the high dimension, the matrix inversion is exactly what needs to be avoided.
To solve the generalized eigenvalue problem with acceptable complexity, we have utilized the conjugate gradient (CG) methods \cite{Yang1993Conjugate}, which approaches the minimum generalized eigenvalue by an iterative method.
The algorithm of the low-complexity framework is illustrated in Algorithm \ref{Alg:LF-Precoding}.
In the rest of this section, we will provide the detailed analysis of the precoder in the two extremes.

\begin{algorithm}[htb]
	\caption{Low-complexity Framework for Robust Precoding}
	\label{Alg:LF-Precoding}
	\footnotesize
	\begin{algorithmic}[1]
		\Require The channel matrices $\bar \bh_k$ and $\bomega_k, k=1,\ldots,K$, the noise variance $\sigma_n$ and total power constraint $P$
		\Ensure The precoding matrices $\hat \bp_k, k=1,\ldots,K$
		
		\State Compute the corresponding parameters $\beta_k, k=1,\ldots,K$.
		
		\State Compute the instantaneous precoding power ${\bm \rho}_\bh$ by (\ref{Equ:I_rho}) and the instantaneous Lagrange multipliers ${\bm \mu}_\bh$ by (\ref{Equ:I_mu}).
		
		\State Compute the statistical Lagrange multipliers ${\bm \mu}_\bomega$ by (\ref{Equ:S_mu}) and the statistical precoding power ${\bm \rho}_\bomega$ by (\ref{Equ:S_rho}).
		
		\State Compute the Lagrange multipliers by (\ref{Equ:W-mu}) and the precoding power by (\ref{Equ:W-rho}).
		Delete users with $\mu_k \le \epsilon$.
		
		\State 
		Compute the normalized precoding vector $\underline\bp_k$ and the parameter $\gamma_k, k=1,\ldots,K$ in (\ref{Equ:MaxEigvalue}) by conjugate gradient method.
		
		\State Compute the precoding vectors $\bp_k = \sqrt{\rho_k} {\underline\bp}_k, k=1,\ldots,K$.
		
	\end{algorithmic}
\end{algorithm}

\subsection{Instantaneous CSI-Based Precoder}

As has been analyzed in \ref{Subsec:GFS}, the Lagrange multipliers can be computed by the WMMSE approach when only instantaneous CSI is available.
Besides, similar to LMNN, we can train a neural network, which takes $\bar\bH = [\bar \bh_1,\ldots,\bar \bh_K] \in {\mathbb C}^{K \times M_t}$ as the input and ${\bm \mu}_\bh\in {\mathbb C}^{K \times 1}$ as the output.
However, due to the high dimension of channel vectors, the complexity of either WMMSE or neural network is not as low as expected.
Thus, to further reduce the complexity without pursuing the optimal solution, the Lagrange multipliers can be computed by some suboptimal precoding vectors such as the RZF precoder, i.e.,
\begin{align}\label{Equ:RZF}
\bp_k^{rzf} = \xi( K \sigma_n^2 \bI + \bar \bH^H \bar \bH )^{-1} \bar \bh_k,
\end{align}
where $\xi$ is a normalization factor.
Thus, the precoding power of the $k$-th user is
\begin{align}\label{Equ:I_rho}
{\bm \rho}_\bh = \diag {\xi^2 \bar \bH ( K \sigma_n^2 \bI + \bar \bH^H \bar \bH )^{-2} \bar \bH^H},
\end{align}
and the normalized precoding vector can be written as ${\underline \bp_k^{rzf}} = \bp_k^{rzf} / \sqrt{[{\bm \rho}_\bh]_k}$.
Besides, denote $\bW = ( K \sigma_n^2 \bI + \bar \bH^H \bar \bH )^{-1}$, the rate of RZF precoding can be expressed as
\begin{align}
\cR_k^{rzf} =  \log  (1 + \xi^2 (r_k^{rzf})^{-1} | \bar\bh_k^H \bW \bar\bh_k |^2 ),
\end{align}
where $r_k^{rzf} = \sigma_n^2 + \xi^2 \sum_{i \neq k} | \bh_k^H \bW \bh_i |^2$.
Denote $\gamma_k^{rzf} = 2^{\cR_k^{rzf}}-1$, similar to (\ref{Equ:Cal_mu}), the Lagrange multipliers can be computed by
\begin{align}\label{Equ:I_mu}
{\bm \mu}_\bh = \sigma_n^2 (\bT_\bh^{-1})^H {\bf 1}_{K \times 1},
\end{align}
where
\begin{align}
[\bT_\bh]_{ki} = 
\begin{cases}
\frac{1}{\gamma_k^{rzf}} | \bar\bh_k^H \underline\bp_i^{rzf} |^2, & k = i, \\
- | \bar\bh_k^H \underline\bp_i^{rzf} |^2,                  & k \neq i.
\end{cases}
\end{align}

\subsection{Statistical CSI-Based Precoding}
As analyzed before, only once computation is required during the period of time-frequency resources.
Thus, it is acceptable to compute the precoding vector by an iterative approach.
However, in some specific communication systems, different subcarriers and slots may be assigned to different users, where the statistical CSI is not same.
To expand the scope of application, we propose to compute the statistical Lagrange multipliers by statistical CSI learning, which is similar to the strategy in the general framework.
To be more specific, we utilize the neural networks to obtain the Lagrange multipliers.
The structure of the statistical Lagrange multipliers neural network (SLMNN) is similar to LMNN, the only difference is that the input of SLMNN is only statistical CSI.
The detailed training progress can be seen in Section \ref{Sec:GF}.
Denote $\bOmega = [\bomega_1,\ldots,\bomega_K]^H \in {\mathbb C}^{K \times NM_t}$, the function of the SLMNN can be expressed as
\begin{align}\label{Equ:S_mu}
{\bm \mu}_\bomega = f_{{\bm \mu}_\bomega} (\bOmega,\nu;\bw_\bomega),
\end{align}
where the set of all weight and bias parameters have been grouped together into a vector $\bw_\bomega$.
Similar to (\ref{Equ:T}), we can compute matrix $\bT_\bomega$ by setting $\beta_k = 0, \forall k$.
Thus, the precoding power can be computed by
\begin{align}\label{Equ:S_rho}
{\bm \rho}_\bomega = \sigma_n^2 \bT_\bomega^{-1} {\bf 1}_{K \times 1}.
\end{align}

\section{Simulation Results}\label{Sec:Simulation}

In this section, we present simulation results to evaluate the performance of the proposed approaches, using the QuaDRiGa channel model \cite{Jaeckel2014QuaDRiGa}, which is a 3-D geometry-based stochastic model with time evolution.
In particular, we consider a massive MIMO system consisting of one BS and $K = 40$ users.
The BS is equipped with $M_t = 128$ antennas (UPA, $M_v = 8$, $M_h =16$) and the height of BS is $25$m.
Users with single antenna are randomly distributed in the cell with radius r = $100$m at $1.5$m height.
Each time slot consists of $10$ blocks, each block takes up $0.5$ms and contains $84$ samples taken from $12$ subcarrires of $7$ orthogonal frequency-division multiplexing (OFDM) symbols.
The center frequency is set at 4.8 GHz.
For the QuaDRiGa model, we consider the  \textit{3GPP\_3D\_UMa\_NLOS} (urban macro) scenario \cite{Jaeckel2014QuaDRiGa} and utilize oversampling DFT matrix (oversampling factor $N_v = 2$, $N_h = 2$) to transform channels into the beam domain.
Three mobile scenarios with moving speeds $30$, $\,80$ and $240$ kmph, are considered.

\subsection{Neural Networks Performance}
The major parameters of neural networks are shown in Table \ref{Tab:MajorPara}.
The input of LMNN can be expressed as
\begin{align}
\bX = [{\rm Re} (\bH_{\bm \beta}), {\rm Im} (\bH_{\bm \beta}), \bOmega_{\bm \beta}]^H.
\end{align}
The dimension of input is and $768 \times 40$ and the size of extracted feature after four convolution modules is $1 \times 40 \times 2$, which can be flattened into a vector $\bf m$.
Furthermore, group $\bf m$ and $\nu$ into a $81 \times 1$ vector as the input of the fully-connected layers, the unit number of hidden layer is $1024$ and the output is $\bm \mu$.
The structure of the SLMNN is similar, the differences are that the input of convolution modules is $\bOmega_{\bm \beta}$ and the hyper-parameters are partially different.
The other main parameters are shown on the right side of the table, which are shared by the two networks.
\begin{table}[htb]
	\scriptsize
	\caption{Major Parameters of Neural Networks} \label{Tab:MajorPara}
	\centering
	\ra{1.3}
	\begin{tabular}{cccccc|lc}
		\toprule
		\multicolumn{3}{c}{LMNN (Input Size: $768 \times 40$)} & \multicolumn{3}{c|}{SLMNN (Input Size: $512 \times 40$)} & \multicolumn{2}{c}{Other Hyper-parameter}\\
		\midrule
		Kernel Size (Num) & Pooling & Feature Size & kernel Size (Num) & Pooling & Feature Size & Dataset Size & 160000 \\
		
		$48 \times 5\;(4)$ & $8 \times 1$ & $96 \times 40 \times 4$
		& $32 \times 5\;(4)$ & $8 \times 1$ & $64 \times 40 \times 4$
		& Batchsize & 1024\\
		
		$24 \times 5\;(8)$ & $6 \times 1$ & $16 \times 40 \times 8$
		& $16 \times 5\;(8)$ & $4 \times 1$ & $16 \times 40 \times 8$ & Algorithm & ADMA\\
		
		$\;\;8 \times 5\;(4)$ & $4 \times 1$ & $\;\;4 \times 40 \times 4$
		& $\;\;8 \times 5\;(4)$ & $4 \times 1$ & $\;\;4 \times 40 \times 4$ & Learning Rate & 0.001\\
		
		$\;\;4 \times 5\;(2)$ & $4 \times 1$ & $\;\;1 \times 40 \times 2$
		& $\;\;4 \times 5\;(2)$ & $4 \times 1$ & $\;\;1 \times 40 \times 2$ & Dropout & 0.5\\
		
		\multicolumn{3}{c}{$81\,-\,1024\,-\,40\,$} & \multicolumn{3}{c|}{$81\,-\,1024\,-\,40$} & Training Steps  & 10000 \\
		\bottomrule
	\end{tabular}
\end{table}

As the dataset is generated off-line, the computational complexity of iterative approach is affordable.
Thus, the number of iterations is set as 20, which is large enough to converge.
Besides, to enhance the generalization performance, various scenarios are considered in dataset, e.g., different mobile velocities, SNRs, user distributions, etc.
As such, the trained neural network can be applied to various practical scenarios.
It is worth mentioning that the iterative algorithm achieves local optimal solutions by optimizing precoding vectors instead of the Lagrange multipliers to maximize the sum rate.
In such a case, the iterative approach is robust and different initial values achieve solutions with similar sum rate, even if the corresponding Lagrange multipliers may differ sometimes.
Table \ref{Tab:Example} shows an example of the above situation, which means the same channel matrices may achieve different Lagrange multipliers due to random initial values.
For these considerations, 10 initial values (including one RZF solution, one SLNR solution, and 8 random values) are iterated, respectively, and the best one is chosen to be one sample for robustness against accidentally bad local optimal solutions. 
\begin{table}[htb]
	\footnotesize
	\caption{An example of Lagrange multipliers and sum rate} \label{Tab:Example}
	\centering
	\ra{1.3}
	\begin{tabular}{cc}
		\toprule
		Lagrange multipliers & sum rate (bit/s/Hz)\\
		$[0.3976,\;0.5054,\;0.4801,\;0,\;0.4821,\;\ldots]$ & 221.9684 \\
		$[0.6659,\;\;\;0,\;\;\;0,\;\;\;0.8371,\;\;\; 0.6224,\;\ldots]$ & 219.6985 \\
		\bottomrule
	\end{tabular}
\end{table}

\begin{figure}
	\centering
	\subfigure[LMNN]{
		\includegraphics[width=3.1 in]{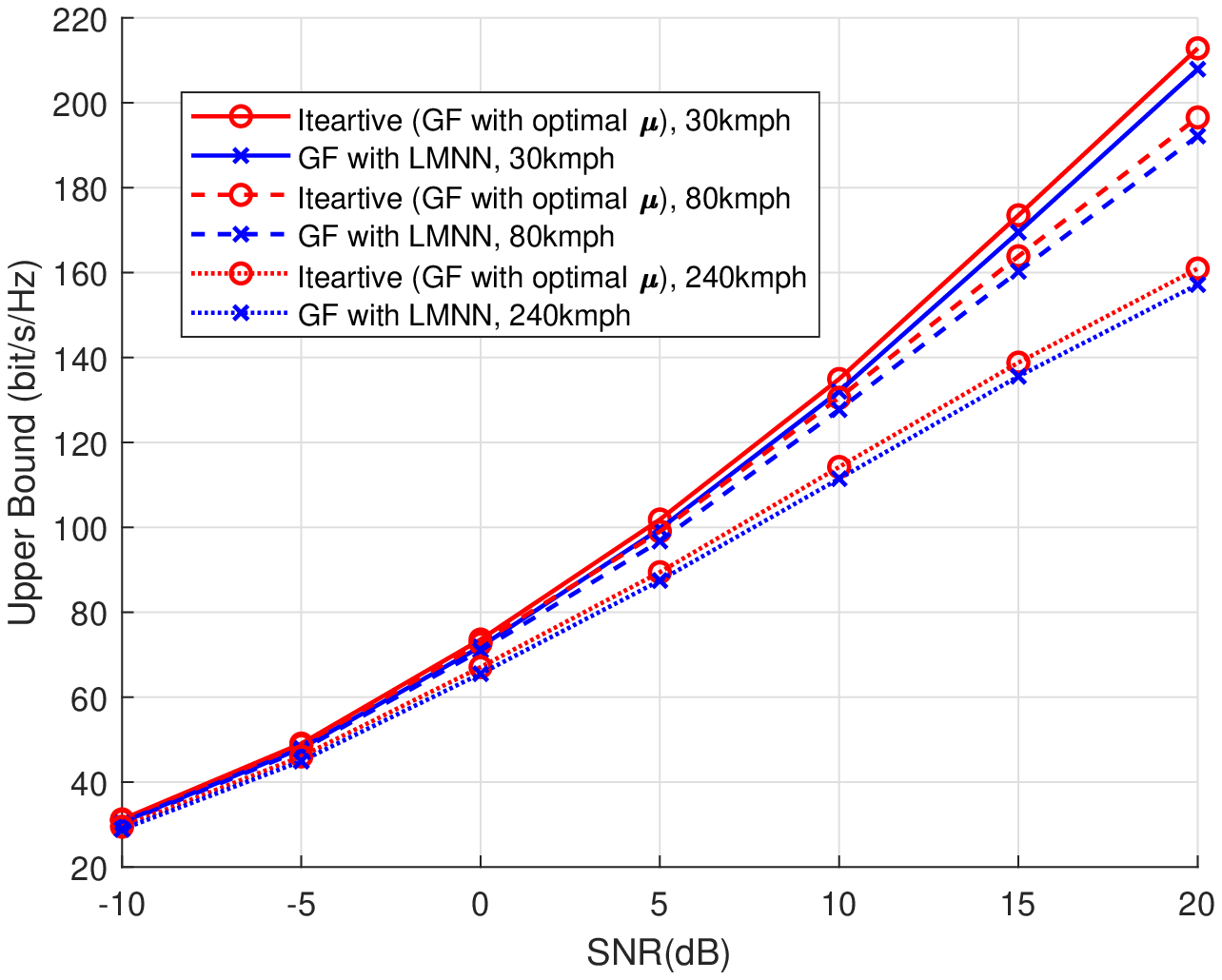}
	}
	\subfigure[SLMNN]{
		\includegraphics[width=3.1 in]{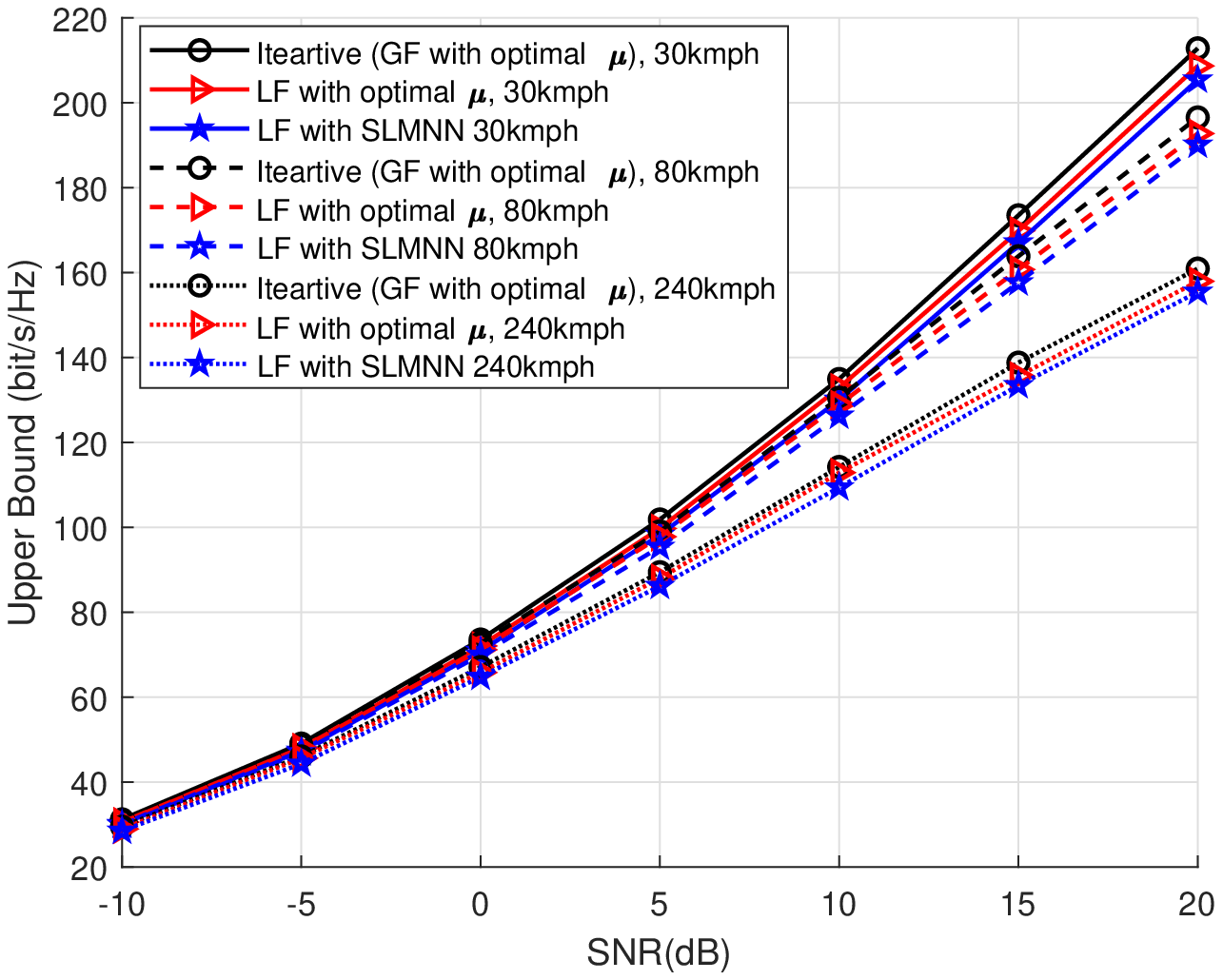}
	}
	\caption{Sum rate upper bound of LMNN-based and SLMNN-based frameworks versus SNR in various mobile scenarios.}\label{Fig:UpperBound}
\end{figure}

To evaluate the performance of the proposed neural networks, we first simulate the upper bound of the  ergodic rate.
Fig. \ref{Fig:UpperBound}$\,$(a) shows the sum rate upper bound of the LMNN-based general framework versus SNR in various mobile scenarios.
Since the data set is generated from the iterative approach in (\ref{Equ:Iterative}), we take it as a benchmark.
As can be seen, the LMNN-based general framework achieves near-optimal performance in various mobile scenarios.
Fig. \ref{Fig:UpperBound}$\,$(b) shows the sum rate upper bound of the SLMNN-based low-complexity framework versus SNR in various mobile scenarios.
The iterative approach and the weighting strategy with the optimal Lagrange multipliers (computed by the solution of iterative approach) are presented here as benchmarks to evaluate the loss of the weighting strategy and performance of the SLMNN, respectively.
There exists a little performance loss in the low-complexity framework due to the weighting operation.
Besides, little gap between the optimal $\bm \mu$ and SLMNN implies the near-optimal performance of the neural networks.

\subsection{Sum Rate Performance of Proposed Frameworks}

\begin{figure}[htb]
	\setlength{\abovecaptionskip}{2pt}
	\setlength{\belowcaptionskip}{8pt}
	\centering
	\includegraphics[width=3.5 in]{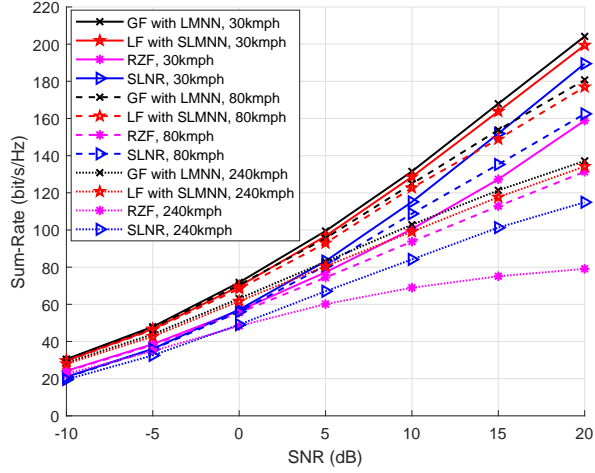}
	\caption{Sum rate versus SNR with respect to different precoding approaches.}\label{Fig:SumRate}
\end{figure}

We further simulate the sum rate to evaluate the performance of the proposed frameworks.
Fig. \ref{Fig:SumRate} shows the sum rate versus SNR with respect to different precoding approaches.
The RZF precoder in (\ref{Equ:RZF}) and the SLNR precoder in (\ref{Equ:SLNR}) are presented here as a baseline.
As can be seen, the RZF precoder works well in the low-mobility scenario.
However, it deteriorates rapidly as the mobile velocity increases.
Besides, the SLNR precoder works better than RZF.
However, the gap between the SLNR precoder and the proposed frameworks grows with the increasing speed.
In the case of 240 kmph at 20 dB, there exists about $19.3\%$ and $73.1\%$ gains of the sum rate in the LMNN-based framework compared with the SLNR and the RZF precoders, respectively.
It is not surprising that the performance of the RZF and SLNR precoders are unsatisfactory as the former takes no advantages of statistical CSI and the latter does not directly maximize the sum rate.
The results show the improved performance of the proposed frameworks, especially in high-mobility scenarios.


\section{Conclusion}\label{Sec:Conclusion}
In this paper, we have proposed a deep learning approach for downlink precoding in massive MIMO, making use of instantaneous and statistical CSI simultaneously.
By transforming the ergodic rate maximization problem into a QoS one, the optimal solution structure is characterized.
With a Lagrangian formulation, the precoding directions and powers can be computed by solving a generalized eigenvalue problem that relies only on available CSI and the Lagrange multipliers.
As such, the high-dimensional precoding design can be alternatively done by low-dimensional Lagrange multipliers, which can be computed by a learning approach.
In particular, a neural network is designed to learn directly the mapping from CSI to the Lagrange multipliers, and then the precoding vectors are computed by solution structure without resorting to iterative algorithms.
To further reduce the computational complexity, we decompose each Lagrange multiplier into two parts, corresponding to instantaneous and statistical CSI, respectively, so that these two parts can be learned separately with reduced complexity.
It is observed from simulation results that the general framework achieves the near-optimal performance and the low-complexity framework greatly reduces the computational complexity but with negligible performance degradation.

\appendices

\section{Proof of Theorem \ref{Lemma:EquivalentProblem}}
\label{Proof:EquivalentProblem}
Denote $\bp_k = \sqrt{\rho_k} {\underline \bp}_k$, where $\rho_k$ is the power allocated to the $k$-th user, ${\underline \bp}_k$ is normalized precoding vector satisfying ${\underline \bp}_k^H{\underline \bp}_k = 1$.
The rate of $k$-th user can be rewritten as
\begin{align}
\cR_k = \mathbb E \big\{\log (\sigma_n^2 \bI + \sum_{i =1}^{K} \rho_i \bh_k^H{\underline \bp}_i{\underline \bp}_i^H\bh_k ) \big\} - \log \mathbb E \big\{ (\sigma_n^2 \bI + \sum_{i \neq k}^{K} \rho_i \bh_k^H{\underline \bp}_i{\underline \bp}_i^H\bh_k ) \big\}.
\end{align}
Then $\bf P2$ can be rewritten as
\begin{equation}
\begin{array}{l}
\mathop {\min }\limits_{\rho_1,\ldots,\rho_K,{\underline \bp}_1,\ldots,{\underline \bp}_k} \sum \limits_{k=1}^K \rho_k, \\
\;\;\;\;\;\;\; {\rm{s.t.}} \;\;\;\;\; \cR_k \ge \cR_k^\Diamond,\\
\;\;\;\;\;\;\;\;\;\;\;\;\;\;\;\;\; {\underline \bp}_k^H{\underline \bp}_k = 1,
\end{array}
\end{equation}
whose optimal solution and corresponding ergodic rates are denoted by $(\rho_1^\star,...,\rho_K^\star,{\underline \bp}_1^\star,...,{\underline \bp}_K^\star)$ and $\cR_1^\star,\ldots,\cR_K^\star$, respectively.

Owing to the constraint $\cR_k \ge \cR_k^\Diamond$,  assume there exists $\cR_m^\star$ satisfying
\begin{equation}\label{Equ:Nothold}
\cR_m^\star > \cR_m^\Diamond.
\end{equation}
It is easy to verify that $\cR_k$ monotonically increases with the power allocated to itself $\rho_k$ and decreases with the power allocated to other user $\rho_i,i \neq k$.
As $\cR_k$ is continuous with respect to $\rho_m$, there always exists a sufficiently small $\varepsilon$ to establish a solution $(\rho_1^\star,...,\rho_m^\star - \varepsilon,...,\rho_K^\star,{\underline \bp}_1^\star,...,{\underline \bp}_K^\star)$ whose corresponding rates $({\hat \cR}_1,...,{\hat \cR}_K)$ satisfy
\begin{equation}
{\hat \cR}_k = 
\begin{cases}
\cR_k^\star - \varepsilon_k > \cR_k^\Diamond, & k = m \\
\cR_k^\star + \varepsilon_k > \cR_k^\Diamond, & k \neq m
\end{cases},
\end{equation}
where variables $\varepsilon_k > 0$ are sufficiently small.
Thus, the solution $(\rho_1^\star,...,\rho_m^\star - \varepsilon,...,\rho_K^\star,{\underline \bp}_1^\star,...,{\underline \bp}_K^\star)$ satisfies the constraint and achieves lower objective, simultaneously.
This is contrary to that $(\rho_1^\star,...,\rho_K^\star,{\underline \bp}_1^\star,...,{\underline \bp}_K^\star)$ is the optimal solution.
As a result, we can obtain that (\ref{Equ:Nothold}) does not hold and
\begin{equation}
\cR_m^\star = \cR_m^\Diamond,
\end{equation}
i.e., $\bf {S2}$ achieves the same ergodic rates as $\bf {S1}$.
In addition, obviously $\bf {S1}$ is a flexible solution for $\bf P2$ so that the optimal solution $\bf {S2}$ achieve lower or equal objective (total power).

When $\bf {S1}$ is global optimal, it achieves the same total power as $\bf {S2}$.
If otherwise, a different solution by increasing the total power of $\bf {S2}$ can achieve a higher objective of $\bf P1$ while still subject to the total power constraint, which contradicts the assumption that $\bf {S1}$ is global optimal.

\section{Proof of Theorem \ref{Theorem:MaxEigvalue}}\label{Proof:MaxEigvalue}
Let $\lambda_k^{[n_k]}$ denote the $n_k$-th largest generalized eigenvalue of matrix pair $(\bS_k,\bN_k)$, we have
\begin{align}\label{Equ:GEVP_k}
\mu_k \bR_k \underline\bp_k^{[n_k]} = \lambda_k^{[n_k]} \Big (\sigma_n^2\bI + \sum_{i \neq k}^{K} \mu_i \bR_i \Big ) \underline\bp_k^{[n_k]}.
\end{align}
Construct the precoding vector $\bp_k^{[n_k]} = \sqrt{\rho_k^{[n_k]}} \underline\bp_k^{[n_k]}$, where $\rho_k^{[n_k]},\forall k$ satisfies the following equations
\begin{align}\label{Equ:Constraint_k}
\sigma_n^2 + \sum_{i \neq k}^{K} \rho_i^{[n_i]} (\underline\bp_i^{[n_i]})^H \bR_k \underline\bp_i^{[n_i]} - \frac{\rho_k^{[n_k]}}{\lambda_k^{[n_k]}} (\underline\bp_k^{[n_k]})^H \bR_k \underline\bp_k^{[n_k]}= 0, \;\; k=1,\ldots,K.
\end{align}
Similar to Lemma \ref{Lemma:Non-singular}, $\rho_k^{[n_k]}$ uniquely exists.
Let (\ref{Equ:GEVP_k}) left-multiplied by $\rho_k^{[n_k]} (\underline\bp_k^{[n_k]})^H$ and let (\ref{Equ:Constraint_k}) left-multiplied by $\mu_k$, then sum up these equations of all users, we have
\begin{align}
&\sum_{k=1}^K (1 + \frac{1}{\lambda_k^{[n_k]}}) \mu_k (\bp_k^{[n_k]})^H \bR_k \bp_k^{[n_k]} = \sum_{k=1}^K \Big (\sigma_n^2\rho_k^{[n_k]} + \sum_{i = 1}^{K} \mu_i (\bp_k^{[n_k]})^H \bR_i \bp_k^{[n_k]} \Big ), \\
&\sum_{k=1}^K (1 + \frac{1}{\lambda_k^{[n_k]}}) \mu_k (\bp_k^{[n_k]})^H \bR_k \bp_k^{[n_k]} = \sum_{k=1}^K \Big (\sigma_n^2\mu_k + \sum_{i = 1}^{K} \mu_k (\bp_i^{[n_i]})^H \bR_k \bp_i^{[n_i]} \Big ).
\end{align}
By combining the results, we have
\begin{align}
\sum_{k=1}^K \mu_k = \sum_{k=1}^K\rho_k^{[n_k]} \le P, \;\; \forall n_k,
\end{align}
where the sign `$\le$' is because that one set of $\{\rho_k^{[n_k]}\}$ is the power of optimal solution.
This means for all $n_k$, $\bp_1^{[n_1]},\ldots,\bp_K^{[n_K]} $ can achieve the minimum power although it may not be flexible.
Besides, from (\ref{Equ:Constraint_k}) we have
\begin{align}\label{Equ:SumRate_Eig}
\cR_k^{ub}(\bp_1^{[n_1]},\ldots,\bp_K^{[n_K]}) = \log(1 + \lambda_k^{[n_k]}).
\end{align}
Denotes by $n_k^\Diamond$  the index of the $k$-th user's optimal eigenvalue.
Assume that $\lambda_k^{[n_k^\Diamond]}$ is not the maximum generalized eigenvalues, then there always exists another eigenvector of a larger eigenvalue, which simultaneously achieves the minimum total power and higher rate, while the rates of other users remain unchanged because of the power control of (\ref{Equ:Constraint_k}).
Similar to Appendix \ref{Proof:EquivalentProblem}, we can reduce the power of this user to achieve lower total power and simultaneously still satisfy the constraints.
This reveals $\bp_1^{[n_1^\Diamond]},\ldots,\bp_K^{[n_K^\Diamond]}$ is not the optimal solution, which is contradictory.
Thus, $\gamma_k$ is the maximum generalized eigenvalue.
This completes the proof.

\section{Proof of Lemma \ref{Lemma:Non-singular}}
\label{Proof:Non-singular}
Denote the matrix $\bQ = \bT \bLambda$, where $\bLambda = \diag{\rho_1,\ldots,\rho_K}$.
According to (\ref{Equ:KKT_C_1}), we have
\begin{align}
\sum\limits_{j \neq k}^K q_{kj} = q_{kk} - \sigma_n^2 < q_{kk}, \;\;  k = 1,\ldots,K,
\end{align}
where $[\bQ]_{ki} = q_{ki}$.
This means the matrix $\bQ$ is strictly diagonally dominant.
Thus, we have that $\bQ$ is non-singular \cite[Theorem 6.1.10 (a)]{Roger2013Matrix}.
As $\bLambda \succ {\bf 0}$ is non-singular, the matrix $\bT = \bQ \bLambda^{-1}$ is non-singular.
This completes the proof.

\section{Proof of Theorem \ref{Theorem:LMO}}\label{Proof:LMO}
Denote $(\mu_1^\Diamond,\ldots,\mu_K^\Diamond)$ the optimal Lagrange multipliers  of $\bf P3$.
As has been proved in Appendix \ref{Proof:MaxEigvalue} that $\sum\nolimits_{k=1}^K \mu_k^\Diamond\le P$, we have $(\mu_1^\Diamond,\ldots,\mu_K^\Diamond)$ is a feasible solution of $\bf P4$.
Besides, $\forall \mu_k,k=1,\ldots,K$ which satisfying $\sum\nolimits_{k=1}^K \mu_k \le P$, a set of precoding vectors $(\bp_1,\ldots,\bp_K)$ satisfying $\sum\nolimits_{k=1}^K \bp_k^H \bp_k  \le P$ can be constructed utilizing the strategy in Appendix \ref{Proof:MaxEigvalue} and the corresponding rate upper bound can be expressed as $\cR^{ub}_k = \log \big(1 + \rho ( \bN_k^{-1} \bS_k) \big) = {\check \cR_k}$.
Assume that $(\mu_1^\Diamond,\ldots,\mu_K^\Diamond)$ is not the optimal solution of $\bf P4$, i.e., existing $(\mu_1^\star,\ldots,\mu_K^\star)$ whose objective function and constructed precoding vectors satisfy
\begin{align}
&f({\check \cR}_1^\star,\ldots,{\check \cR}_K^\star) > f({\check \cR}_1^\Diamond,\ldots,{\check \cR}_K^\Diamond), \\
&\sum\limits_{k=1}^K (\bp_k^\star)^H \bp_k^\star = \sum\limits_{k=1}^K \mu_k^\star \le P.
\end{align}
As $\bf S1$ is global optimal, noting that $\bf P1$, $\bf P2$ and $\bf P3$ are equivalent when employing the upper bound simultaneously.
This means $(\bp_1^\Diamond,\ldots,\bp_K^\Diamond)$ (constructed by $\mu_k^\Diamond$) is not the optimal solution of $\bf P1$, i.e., not the optimal solution of $\bf P3$, which is contradictory.
Thus, $(\mu_1^\Diamond,\ldots,\mu_K^\Diamond)$ is the optimal solution of $\bf P4$.
This completes the proof.

\bibliographystyle{IEEEtran}
\bibliography{IEEEabrv,IEEEConf}

\end{document}